\documentclass[sigconf,authorversion]{acmart}

\usepackage{caption}
\usepackage{subcaption}
\usepackage{multirow}
\usepackage{algorithm2e}
\usepackage{breqn}

\AtBeginDocument{%
  \providecommand\BibTeX{{%
    \normalfont B\kern-0.5em{\scshape i\kern-0.25em b}\kern-0.8em\TeX}}}

\copyrightyear{2024} 
\acmYear{2024} 
\setcopyright{acmlicensed}
\acmConference[IUI '24]{29th International Conference on Intelligent User Interfaces}{March 18--21, 2024}{Greenville, SC, USA}
\acmBooktitle{29th International Conference on Intelligent User Interfaces (IUI '24), March 18--21, 2024, Greenville, SC, USA}
\acmDOI{10.1145/3640543.3645152}
\acmISBN{979-8-4007-0508-3/24/03}

\begin{document}

\title[An Incremental Learning Multimodal Object Referencing Framework]{Looking for a better fit? An Incremental Learning Multimodal Object Referencing Framework adapting to Individual Drivers}

\author{Amr Gomaa}
\orcid{0000-0003-0955-3181}
\affiliation{%
  \institution{DFKI, Saarland Informatics Campus}
    \city{Saarbr{\"u}cken}
  \country{Germany}
}
\email{amr.gomaa@dfki.de}

\author{Guillermo Reyes}
\orcid{0000-0003-4064-8605}
\affiliation{%
  \institution{DFKI, Saarland Informatics Campus}
    \city{Saarbr{\"u}cken}
  \country{Germany}
}
\email{guillermo.reyes@dfki.de}

\author{Michael Feld}
\orcid{0000-0001-6755-5287}
\affiliation{%
  \institution{DFKI, Saarland Informatics Campus}
    \city{Saarbr{\"u}cken}
  \country{Germany}
}
\email{michael.feld@dfki.de}

\author{Antonio Kr{\"u}ger}
\orcid{0000-0002-8055-8367}
\affiliation{%
  \institution{DFKI, Saarland Informatics Campus}
    \city{Saarbr{\"u}cken}
  \country{Germany}
}

\email{antonio.krueger@dfki.de}

\renewcommand{\shortauthors}{Gomaa et al.}

\begin{abstract}

The rapid advancement of the automotive industry towards automated and semi-automated vehicles has rendered traditional methods of vehicle interaction, such as touch-based and voice command systems, inadequate for a widening range of non-driving related tasks, such as referencing objects outside of the vehicle. Consequently, research has shifted toward gestural input (e.g., hand, gaze, and head pose gestures) as a more suitable mode of interaction during driving. However, due to the dynamic nature of driving and individual variation, there are significant differences in drivers' gestural input performance. While, in theory, this inherent variability could be moderated by substantial data-driven machine learning models, prevalent methodologies lean towards constrained, single-instance trained models for object referencing. These models show a limited capacity to continuously adapt to the divergent behaviors of individual drivers and the variety of driving scenarios. 
To address this, we propose \textit{IcRegress}, a novel regression-based incremental learning approach that adapts to changing behavior and the unique characteristics of drivers engaged in the dual task of driving and referencing objects. We suggest a more personalized and adaptable solution for multimodal gestural interfaces, employing continuous lifelong learning to enhance driver experience, safety, and convenience. Our approach was evaluated using an outside-the-vehicle object referencing use case, highlighting the superiority of the incremental learning models adapted over a single trained model across various driver traits such as handedness, driving experience, and numerous driving conditions. Finally, to facilitate reproducibility, ease deployment, and promote further research, we offer our approach as an open-source framework at \url{https://github.com/amrgomaaelhady/IcRegress}.

\end{abstract}



\begin{teaserfigure}
  \includegraphics[width=\linewidth]{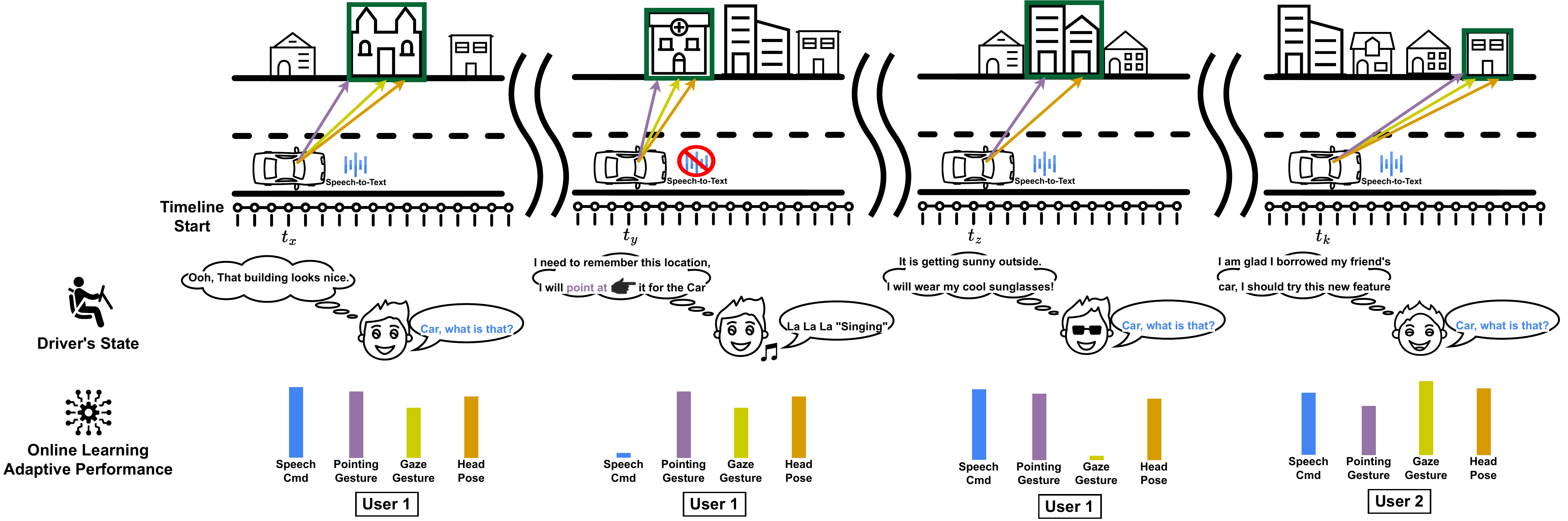}
  \caption{
  An illustrative example of the incremental learning approach for multimodal object referencing that adapts to various drivers and driver states. The user references a building by pointing and looking at it and inquires about it from the vehicle. The model undergoes continuous adaptation through incremental learning online. Training commences on the left with all available modalities (an ideal setting) to establish a base model, which is further adapted for scenarios where specific modalities are unavailable or a different driver is operating the system.
  }
  \Description{An illustration of a user driving his car while asking the car about a referenced object by pointing, looking at it, and saying the speech command ``What is that?''. The figure shows four different scenarios to reference objects. From left to right, they start with an ideal case showing that all the modalities are available. Then, it shows that the speech command is missing, as the driver is listening to loud music, but the model has adapted to that and still correctly identified the building without needing the speech command. In the following scenario, the gaze data is missing because the driver decided to wear sunglasses. Furthermore, the final scenario shows that another driver is using the object referencing system, and the model has to adapt the tracked modality for the new driver. There is a timeline to indicate that the process is continuous.}
  \label{fig:teaser}
\end{teaserfigure}


\maketitle

\section{Introduction}

With significant progress in the development of autonomous vehicle technology, there has been a growing interest in exploring non-traditional approaches to enhance human-vehicle interaction inside the vehicle using the available sensors such as gaze and hand trackers. Therefore, researchers and practitioners have been actively exploring the concept of intelligent interiors to improve user experience, acceptance, and trust in AV technology~\cite{murali2022intelligent}. Moreover, this allowed researchers to introduce novel use cases for vehicle interaction, including outside-the-vehicle object referencing. Although vehicular external imaging systems such as LiDAR and RADAR allow object detection in the surrounding environment~\cite{wu2020deep,montiel2022towards}, they cannot solve the problem of user-intended object referencing without additional information such as gaze or pointing direction, using vehicle internal sensors. Furthermore, recent work shows that each individual behaves quite differently when performing this referencing task~\cite{gomaa2021ml,Gomaa2020,aftab2021multimodal}. Although existing outside-the-vehicle object referencing approaches have achieved adequate performance~\cite{gomaa2021ml,Gomaa2020,rumelin2013free,2019BMW2019,2021Mercedes-Benz2021}, they are still primarily rigid, one-size-fits-all approaches that are neither adaptable nor incremental and assume that models will be trained once and apply to all driving situations and drivers' states. 

A naive approach for solving the previously mentioned problems is to collect data for all possible driving situations and drivers' states to train a model once (i.e., without incremental learning) and assume that it would have external validity when implemented in a natural system. However, this solution is neither applicable nor feasible as it is impossible to cover all possible driving situations in a highly dynamic and ever-changing environment like driving. It also requires massive data collection and storage systems to store and train this general model. Additionally, this model would average over all these situations and have sub-par performance in all of them due to the critical issue of catastrophic forgetting, among other problems of this architecture~\cite{goodfellow2013empirical,kirkpatrick2017overcoming,de2021continual}. In more detail, ``\textit{Catastrophic Forgetting}'' (from its name) is the problem of losing the prediction power on old tasks when the model attempts to learn new ones; hence, the model \textit{forgets} what it has learned in past information. While one solution for this problem would be to train a different separate model for each driving situation in a many-models-fits-all approach, which would be more adaptable as introduced in~\cite{gomaa2021ml,gomaa2022adaptive}, they still suffer from the same applicability and feasibility issues such as covering all possible driving states, data collection, and storage. Thus, we propose a continual (i.e., incremental) learning approach to overcome these problems, as it only stores the needed information instead of the entire previously trained data, and it constantly learns on the new tasks (i.e., driving situations) to improve the prediction performance of the referenced object.

In conclusion, in this work, we focus on the task of disambiguating the driver's referenced object through the multimodal fusion of gaze, head pose, and pointing while adapting to different drivers, driving scenarios, and driver's specific state as seen in~\autoref{fig:teaser}. We achieve this goal by utilizing incremental learning algorithms commonly used to adapt the learning model for the new emerging task while not ``forgetting'' the previously learned weights corresponding to the base (i.e., old) task. We conducted a user study in a driving simulation environment to collect the needed data and expand the previous work discussion on individual differences when performing this task. \textbf{Our contribution can be summarized in three folds as follows:}

\begin{enumerate}

    \item We propose \textit{IcRegress}, a novel incremental learning approach for regression problems that adapts to different drivers, driver states, and available resources (e.g., sensor availability). We assess the performance of the approach against two baselines with traditional machine learning techniques.
    \item Compared to State-of-the-Art (SOTA) approaches, we introduce new practical metrics for measuring the performance of outside-the-vehicle objects referencing inside a driving simulation captured with existing sensors in modern vehicles and directly transferable to a real car scenario with better scalability and generalization.
    \item We perform an ablation study for different modalities and highlight the importance of each modality to the driver's performance of the referencing task.
\end{enumerate}

\section{Related Work}

Since this work overlaps multiple domains, such as multimodal object referencing in vehicles, personalized model adaptation, and incremental learning algorithms, we highlight essential aspects and state-of-the-art approaches in each domain as follows.

\subsection{Object Referencing in Automotive}

Several researchers studied the interaction with objects within and outside the vehicle using gestures of pointing and gaze~\cite{Ahmad2018a,Fariman2016,Roider2017,feld2016combine,poitschke2011gaze,rumelin2013free,fujimura2013driver,kang2015you,kim2014identification,misu2014situated}.
Fujimura et al.~\cite{fujimura2013driver} investigated the reference of objects outside the vehicle in a simulation environment using hand-pointing gestures. They suggested using constrained-hand pointing instead of free-hand pointing to decrease the risk of the driver's hand being taken off the wheel. However, constrained-hand pointing introduces many limitations on the accurate tracking of the pointing vector, and they did not report exact tracking accuracy figures due to unrealistic approximations.
Consequently, Rümelin et al.~\cite{rumelin2013free} investigated the free-hand pointing approach for object referencing using a lab study with a stationary car and street scenes presented on multiple projectors and a field study using a Wizard-of-Oz technique to collect qualitative feedback.
Additionally, several studies were conducted to monitor a driver's activity using head pose and eye gaze tracking~\cite{ji2002real, ohn2014head, vicente2015driver, vasli2016driver, vora2017generalizing}; however, few focused on the object referencing task~\cite{poitschke2011gaze, kang2015you} using the gaze modality.
Kang et al.~\cite{kang2015you} utilized head pose and eye gaze for referencing objects outside the vehicle using a depth camera in a field study. Similar to referencing using pointing gestures, they considered only the horizontal angles for the referencing task. The estimated referencing angle was the summation of car orientation, head pose, and eye gaze angle. Due to the camera's position (behind the steering wheel), the gaze detection suffered greatly.
Similarly, Poitschke et al.~\cite{poitschke2011gaze} studied in-vehicle object selection and compared it with traditional touchscreen interaction. They utilized a button attached to the steering wheel to determine the onset of the selection task. They showed a significant increase in the selection speed with their approach compared to the touchscreen one. However, they also showed that the driver's cognitive load significantly increased during this shorter period. This load could be alleviated with a more natural interaction using incremental learning approaches that adapt to different situations. Alternatively, Aftab et al.~\cite{aftab2021multimodal} proposed a machine learning-based fusion approach for multimodal object referencing using pointing, gaze, and head pose from a stationary vehicle. In contrast, Gomaa et al.~\cite{Gomaa2020} considered a similar approach from a moving vehicle. However, neither approach considers the varying driving situations and driver traits and provides a one-time model training approach that is non-incremental, unlike in this work.

\subsection{Adaptive Multimodal Interaction}

Adaptive multimodal interaction combining speech, hand gestures, and gaze has been a topic of interest for the research community for the last 20 years in multiple domains, including robotics and automotive applications~\cite{rogers2000adaptive,hassel2005adaptation,janarthanam2014adaptive,manawadu2017multimodal,zhang2015costs,neverova2015moddrop,gnjatovic2012adaptive}.
Researchers attempted multimodal fusion approaches for in-vehicle object selection in multiple works~\cite{roider2018see,Aftab2020,sezgin2009multimodal}. However, in-vehicle object referencing approaches do not generalize directly to outside-the-vehicle referencing, as the object's environment is static, limited, and in close proximity. Consequently, Moniri et al.~\cite{Moniri2012a} studied the single task of outside-the-vehicle object referencing using pointing, head pose, and eye gaze from the passenger seat. Similarly, Aftab et al.~\cite{aftab2021multimodal} combined these modalities using a neural network-based late fusion approach to reference from a stationary vehicle. While both approaches showed great promise, they still considered a stationary single-task situation. As far as we know, there is little work on multimodal fusion for the object-referencing task while driving (i.e., in a dual-task scenario), as most work assumes this task to be used only in fully autonomous vehicles. Gomaa et al.~\cite{gomaa2021ml} proposed a framework for multimodal object referencing in a dual-task scenario without assuming a conditional or fully autonomous setting. They also propose a simple approach to model adaptation based on the Support Vector Machine (SVM) algorithm~\cite{hearst1998support}. However, they utilize that only for personalization and creating a separate model for each driver, which is still lacking in terms of applicability and feasibility, as discussed earlier. Moreover, their approach is specific to support vector regression; thus, it cannot be expanded to other machine learning models, unlike this work, where we suggest an incremental learning algorithm agnostic to the underlying machine learning regression model.

\subsection{Incremental Learning Approaches}

Machine learning (Specifically Deep Learning) approaches have shown remarkable performance in multiple tasks, including object detection and multimodal fusion. However, despite these impressive results, existing approaches mainly consider rigid, static models that are trained once without adapting to the ever-changing traits of human behavior. Thus, continuous learning approaches (e.g., incremental and online learning methods) have been investigated to create adaptive models that learn continuously. Incremental learning approaches are concerned with overcoming the ``Catastrophic Forgetting'' problem~\cite{de2021continual,van2019three,mirza2022efficient,gomaa2023towards,reyes2023personalized} in static models to produce an efficient model that does not require access to a large amount of data and can learn and predict in real-time. Ven et al.~\cite{van2019three} define three scenarios for incremental learning: task-incremental learning, class-incremental learning, and domain-incremental learning. De Lange et al.~\cite{de2021continual} and Zhou et al.~\cite{zhou2023deep} survey existing approaches for class-incremental learning and demonstrate some examples for the other two scenarios. Although existing approaches are quite efficient for classification problems, they are not directly applicable to a regression problem like the one presented in this work. Therefore, we draw inspiration from these approaches to propose a novel approach for incremental learning in regression problems. More specifically, in IcRegress, we adjust the iCaRL~\cite{rebuffi2017icarl} algorithm to a regression task to predict the referencing angle and determine the referenced object in our automotive use case. IcRegress algorithm can be correlated to an intersection between class- and domain-incremental learning. Although IcRegress is evaluated for the referencing object use case, it is a general algorithm that can be utilized for any regression problem.

\begin{figure}[t]
	\begin{center}
		\includegraphics[width=\linewidth]{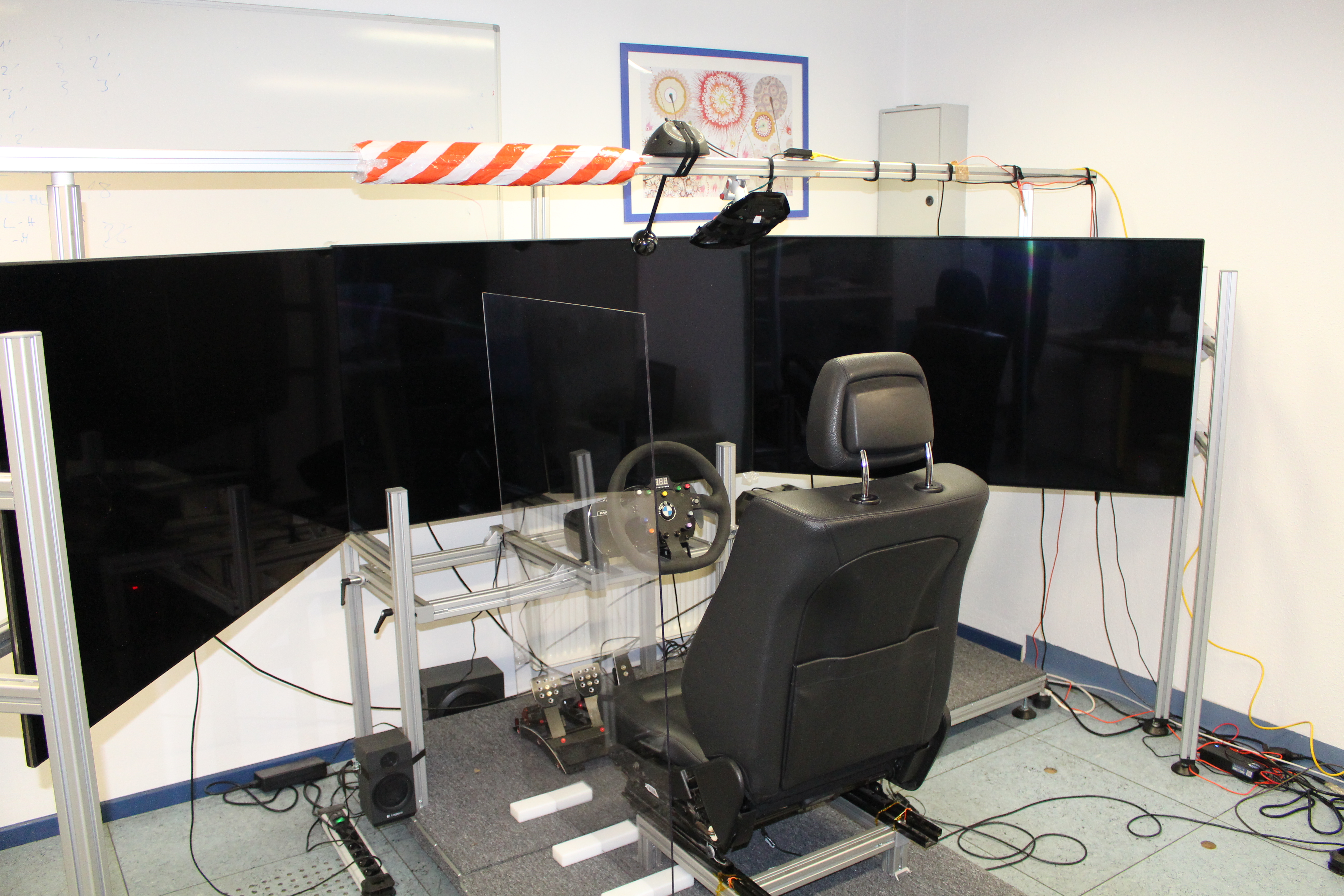}
	\end{center}
	\caption{Setup overview showing our driving simulation with three 55-inch screens, steering wheels, and pedals. We simulate the car's left door with a plastic barrier beside the driving seat. Sensor cameras are attached to a top beam to simulate their location on the roof of modern vehicles.}
    \Description{An image from a lab room shows a driver's seat with driving wheels and pedals. It is surrounded by three 55-inch screens in a semi-circular shape to simulate a vehicle's cockpit. It shows two sensors attached to a metal beam above the driver's seat for gaze and hand tracking.}
	\label{fig:setupSimulator}
\end{figure}

\section{User Study}

In this experiment, we conducted a  counterbalanced within-subjects study in an open-source medium-fidelity driving simulator~\cite{math2013opends}. While the choice of a driving simulation affects the external validity compared to a real car setting, it would focus more on internal validity. This choice comes from the following reasons. First, it gives more control over the environment; for example, we could specify exactly which object the driver would point at, and all users will have the same perspective of this object. Secondly, it is more cost-effective and less risky for the participants. Third, it allows us to record driving behavior precisely and identically recreate all situations for further detailed study in 3D analysis software. Fourth, we can exclude a category of errors stemming from faulty positioning hardware or maps. Finally, since this work focuses on the incremental learning algorithm, we argue that driving simulation will not affect the reproducibility of the algorithm compared to a real driving scenario, as seen in previous work~\cite{murali2022intelligent,montiel2022towards}.

\subsection{Apparatus}
\label{sect:apparatus}

The driving simulator is in an enclosed lab room to ensure the participant is not disturbed during the driving task. During the experiment, the participant sits in a driver's seat surrounded by three 55-inch LCD screens. The steering wheel is mounted on an aluminum stand in front of the participant, and the pedals are below it (see~\autoref{fig:setupSimulator}). The simulation's computer specifications are Intel Xeon Processor E5-1650 v4 3.7 GHz with 128 GB RAM and GEFORCE GTX 1080Ti Graphics Card with 11 GB GDDR5X running on Windows 10. The computer is connected to three 55-inch flat LCD screens configured as one large screen to simulate the entire windshield of a real driving car and the left and right windows of the front seat. We use Fanatec's ClubSport Steering Wheel BMW GT2, which supports force feedback while driving for maximized immersion, and Fanatec's ClubSport Pedals V3. 
Driver hand gestures (i.e., pointing) were captured with a prototype Time-of-Flight (ToF) 3D camera (under a non-disclosure agreement) that is similar to the sensors existing in modern cars.
The driver's face (i.e., head poses and gaze gestures) was captured with a standard webcam and analyzed using the RT-GENE~\cite{fischer2018rt} framework. The driver's speech was captured with an on-body microphone and analyzed using the OpenSmile~\cite{eyben2010opensmile} framework. The voice notification of the referenced object was given using two speakers located 50 cm to the left and right of the steering wheel. Therefore, a preprocessing step was applied to the recorded data before obtaining the final dataset to synchronize and downsample it to a frame rate of 20 Hz (i.e., the frame rate of the slowest device) for all recording devices.

\subsection{Participants}

For this experiment, 59 participants were recruited. However, two did not complete the experiment due to motion sickness and nausea, and one was excluded due to a power outage. Thus, the total number of participants was 56. The gender ratio was almost balanced (29 females and one diverse) with a mean age of 23.71 years (SD = 3.93). Most of the participants were (50 participants) right-handed. Participants were mainly experienced drivers with an average of 5.51 years of driving (SD = 3.73). These different attributes of the driver sample will be used in the incremental learning procedure to produce models adapted to these individual differences.

\subsection{Design}

\begin{figure}[t]
	\begin{center}
		\includegraphics[width=\linewidth]{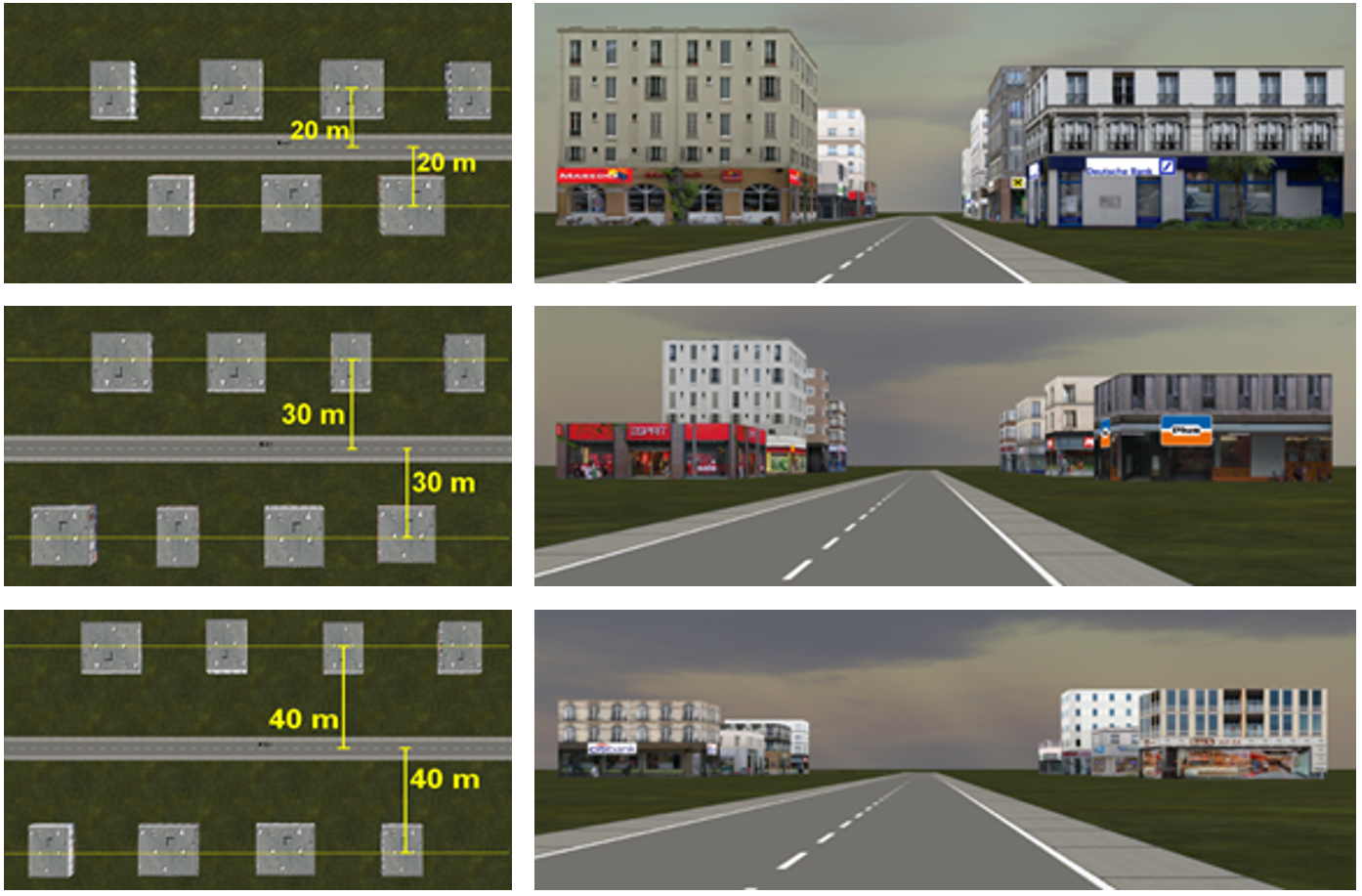}
	\end{center}
	\caption{Buildings' clusters with the three possible lateral building offsets. The Left is the top view, and the right is the driver's view.}
    \Description{Six scenes from OpenDS simulation. The three on the right show a top view for three possible distances from the road for the 8-building clusters. It shows 20, 30, and 40 meters from top to bottom. The other three scenes on the left show the driver's view for the corresponding distance from the road cases.}
	\label{fig:buildingexample}
\end{figure}

\begin{figure}[t]
	\begin{center}
		\includegraphics[width=\linewidth]{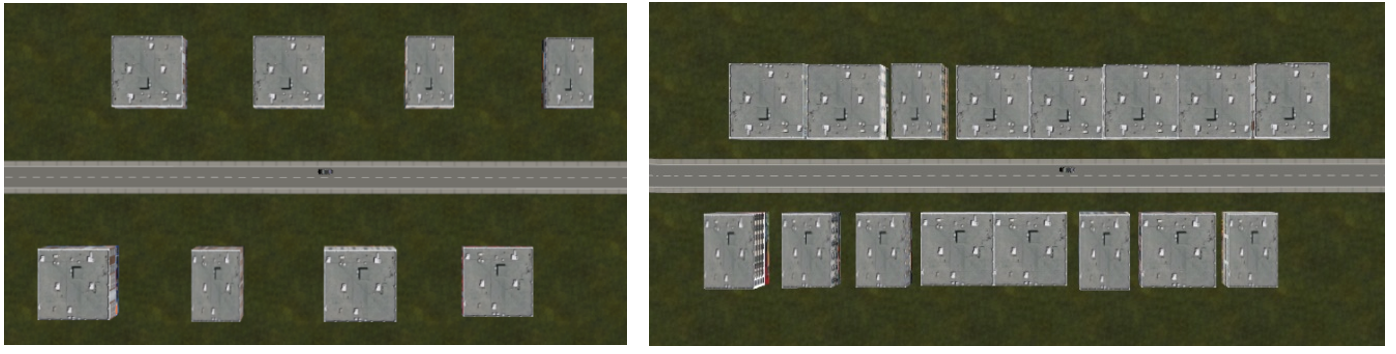}
	\end{center}
	\caption{A top view of 8-building and 16-building clusters.}
    \Description{Two top-view scenes from the OpenDS simulator. The left scene shows building next to each other with a wide gap between them in the 8-building cluster. The right scene shows buildings adjacent to each other with almost no gap between them to indicate the 16-building cluster array. In both scenes, half the buildings stand on the right side of the road, and the other half are on the left.}
	\label{fig:buildingclusterexample}
\end{figure}

\begin{figure}[t]
	\begin{center}
		\includegraphics[width=\linewidth]{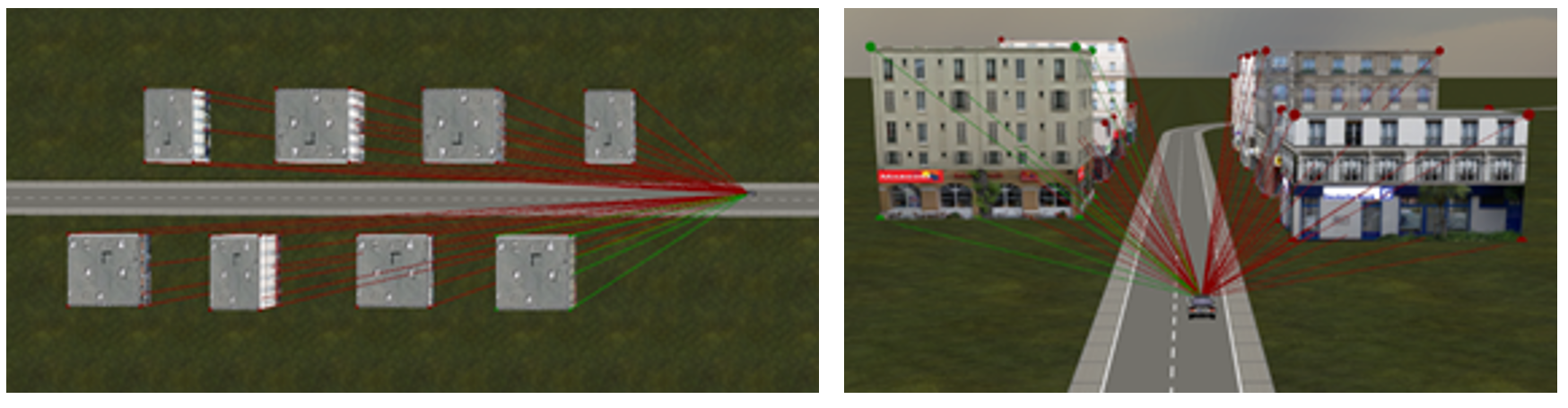}
	\end{center}
	\caption{Cluster with target building visualized using ray casting in OpenDS simulator (top view and lateral view).}
 \Description{Two scenes of an 8-building cluster from the OpenDS simulator. The left scene shows a top-view with ray casting for all the possible buildings in red and the target building in green, while the right scene shows the same from a lateral view showing the driver's car and the same rays.}
	\label{fig:buildingexamplesideview}
\end{figure}

The main section of the experiment consists of the dual task of driving and pointing at buildings. Participants drove one lap on a 49-km long two-lane closed road consisting of 120 straight segments connected by curved segments with an average driving speed of 50 km/h (as per the instructions). For comparability, the instructed driving speed and road design are mapped to related work such as~\cite{gomaa2021ml,Gomaa2020}. The referencing task is performed during straight segments. Those segments had different randomized lengths ranging from 250 to 450 meters to reduce any learning effect, counterbalance data, and increase driver engagement. Similarly, the curved segments vary in direction (left vs. right), angle (45 vs. 90 degrees), and intensity of the curvature (sharp vs. moderate) to reduce the level of monotony and make the driving task more engaging. Buildings of different shapes and sizes were placed along the straight segments on both sides of the road. 
The buildings were placed 20, 30, or 40 meters from the road as seen in~\autoref{fig:buildingexample}. Furthermore, buildings were clustered into sets of 8 and 16 buildings to simulate low and high distracting environments, respectively, as seen in~\autoref{fig:buildingclusterexample}. Finally, the building groups were randomized and counterbalanced to mitigate any confounding factors among the participants.
As for the secondary non-driving task (e.g., referencing task), participants were instructed to perform two secondary tasks, which we classify into \textit{referencing} and \textit{non-referencing}. The reasoning behind having an additional secondary task is to avoid a learning effect on the referencing task and reduce the monotonicity of the task, as suggested in the pilot study and as observed in related work. Since \textit{referencing} is the main interest of this research work, it constituted 80\% of the driving route. In contrast, the \textit{non-referencing} task constituted the remaining segments (i.e., 96 segments were \textit{referencing} tasks, and 24 segments were \textit{non-referencing} ones). Both tasks were randomly distributed among the driving routes for counterbalancing. In the referencing task, exactly one building was chosen as the target building (i.e., Point-of-Interest (PoI)), which should be identified and referenced (i.e., selected) by the driver. In contrast, the remaining seven or fifteen buildings acted as distractors. The distribution of target building was equally randomized among all referencing tasks per participant. For example,~\autoref{fig:buildingexamplesideview} depicts a cluster where the first building on the left has been chosen as the target. 

\subsection{Procedure}

First, participants are greeted at the lab and perform a training driving task to get used to the driving simulator and the secondary tasks (i.e., referencing and non-referencing). Then, the participants started the main route. Although participants were instructed to drive at 50 km/h, they could also drive slower and faster (up to a technical limit of 60 km/h). Therefore, the primary driving time was around 50 to 70 minutes. In the middle of the course, there is a break position where the simulation is paused, and the driver may relax before continuing. When entering a straight-road segment, an audio file's playback will be triggered to instruct the driver verbally (through a pre-recorded automated voice) which building they should reference. The participant was instructed to look for the target building and point at it with the left or right hand while using the verbal instruction ``There it is''. This referencing task produced four distinct modalities for multimodal fusion: pointing, gaze, head pose, and speech. Regarding the non-referencing task, participants were instructed to wave at a pedestrian on the road and say the word ``Hello''. This instruction is also verbal (through a pre-recorded automated voice) and was played by a trigger at the beginning of the respective straight segment, as in the referencing task. In this task, none of the buildings will be considered a target; instead, a pedestrian will walk along the right-hand side of the road toward the driver to resemble the friend at whom he is waving. Before conducting the study, internal and external ethics review boards approved it.

\section{Methodology}

As the main focus of this work is on incremental learning, an essential aspect of the method is the acquired data-splitting strategy. Additionally, the existing evaluation methods for the object referencing task still lack scalability and generalization. Thus, we propose additional metrics to assess the model performance and compare it with existing methods previously used in SOTA approaches (e.g., in~\cite{gomaa2021ml,aftab2021multimodal}). Moreover, we thoroughly describe the machine learning model used for analysis and its input features (i.e., the different tracked modalities).

\subsection{Dataset Split}

The entire data set consisting of N = 56 participants is split into training, validation, and test sets using a 10\% holdout set for validation and a 10\% (i.e., 6 participants) holdout set for testing.
To ensure external validity and model generalization, the test split is done on the participant level, not the data point level. In this way, no data from the same participant is used in training and testing for any trained models. This also conforms to existing data sets and machine learning challenges where the test set is inaccessible to developers during training~\cite{mirza2022efficient}. To evaluate the incremental learning approach, the training and validation sets are further split based on drivers' traits and the availability of modalities to assess different driving settings. These additional splits are based on driving experience (Amateur vs. Expert drivers), handedness (Left handed vs. Right handed), and Speech command availability (Command vs. No command). Driving experience is determined by considering drivers based on the lower and upper quartiles of the participants' distribution, below four years and above six years of driving experience, respectively. Instead of dividing the participants into two halves, choosing the lower and upper quartiles ensures a difference in performance due to the driving experience. Although the test set was randomly chosen from the list of participants, it was equally balanced in driver traits (e.g., three amateur drivers vs. three expert ones).

\subsection{Input features}

The input features of the data set are the pointing, gaze, and head-pose vectors in the X- and Z-directions only (i.e., the horizontal referencing). This is consistent with the previous work that excluded the Y direction (i.e., vertical reference) since the height dimension of PoIs (e.g., the height of buildings) is often irrelevant to this task. As the head pose and the gaze modalities are dependent and often complementary, the gaze and head vectors were summed up here to create an additional vector, which we hereafter call GazeHead. This also conforms to previous work such as~\cite{Gomaa2020,gomaa2021ml,kang2015you}. Thus, the system contains eight input features: the x- and z-coordinates for the pointing, gaze, head-pose, and GazeHead vectors. As mentioned in Section~\ref{sect:apparatus}, all modalities are synchronized to 20Hz. However, they are further downsampled to 5Hz to reduce the feature space size and avoid the curse of dimensionality issue that accompanies data sets with small sizes~\cite{koppen2000curse}. Similarly to related work~\cite{gomaa2021ml,aftab2021multimodal}, we consider a 4-second window around the onset of the referencing (two seconds before and after), resulting in a total of 20 samples per referencing task. The choice of a 4-second window was confirmed by visual inspection of random samples of drivers' referencing. We also compared against a 1-second window as in~\cite{aftab2021multimodal}; however, the resulting input features lacked any relevant information (i.e., the reference information was unavailable within this window). The speech command is a binary trigger that determines this referencing onset. Equation~\ref{eqn:anglepred} shows how the input modalities are used to predict the referencing angle, where $f_{estimator}$ is the regression model applied to the different input modalities for all timestamps from $t_o$ to $t_x$ ($x=20$ for a 4-second window at 5Hz sampling rate) in order to predict the PoI referenced angle at the referencing onset frame. 

\begin{dmath} 
PredictedAngle|_{t=\frac{t_x-t_0 }{2}} = \\ f_{estimator}\left((Pnt_t, GazeHead_t, Gaze_t, Head_t)|_{t=[t_0,...,t_x]}\right)
\label{eqn:anglepred}
\end{dmath}

Thus, the final input of the model is 4 (modalities), $\times$ 2 (coordinates), $\times$ 5 (samples per second), $\times$ 4 (seconds), which is equal to 160 features per data point. The entire data set consists of 5,376 data points (56 participants $\times$ 96 referencing gestures per participant).

\subsection{The Base Learning Model}
\label{sec:baselearningmodel}

Although our incremental learning algorithm is model-invariant as it adapts the input strategy, we still require a machine learning model to train the data. Since the input features consist of time-series data, it requires a learning model that can learn temporal dependency; thus, both a convolutional neural network (CNN) and a long short-term memory (LSTM)~\cite{bishop2006pattern,Goodfellow-et-al-2016} architecture would apply to our data structure. Similar to previous work such as~\cite{gomaa2021ml,aftab2021multimodal}, we chose a convolutional neural network (CNN) approach to train and predict the referenced angle with a similar architecture. For a short time sequence (as in our case), CNN gives similar results to LSTMs with less computational time (for similar parameter size) as seen in~\cite{gomaa2021ml,weytjens2020process}. 
We consider this CNN model to be a back-end deep learning model on which the incremental learning algorithm is applied. The trained model will be called hereafter the ``\textit{Base Model}'' in further methods and results.
We use PyTorch\footnote{\url{https://pytorch.org/}} to implement the CNN architecture. Due to the small size of the training dataset, we use a small number of layers for neural network implementation to reduce the number of learnable parameters; the 1D CNN network consists of three hidden convolutional layers of 64, 16, and 8 feature map sizes (i.e., filters). Each layer is followed by a batch normalization layer for regularization~\cite{santurkar2018does}, a rectified linear unit (ReLU) activation function to add non-linearity, a 1D max pooling layer to reduce the latent vector dimension, and a dropout layer (with p=0.3)~\cite{baldi2013understanding} to reduce overfitting and for further regularization. Finally, since CNN layers are used for feature extraction, they are followed by three fully connected networks to learn the regression task and output the predicted reference angle. The loss function that the network optimizes is the mean square error (MSE), which is typically used for regression problems. The ground truth of the training loss is the angle of the geometric center of the target PoI (i.e., building) at the referencing onset time as in~\cite{gomaa2021ml,aftab2021multimodal}. The network architecture code and data set are available at \url{https://github.com/amrgomaaelhady/IcRegress}. 

\RestyleAlgo{ruled}

\begin{algorithm}[t]
    \SetKwFunction{train}{train}
    \SetKwFunction{inference}{inference}
    \SetKwFunction{rank}{rank}
    \SetKwFunction{retrieveExemplar}{retrieveExemplar}
    \SetKwInOut{KwIn}{Input}
    \SetKwInOut{KwOut}{Output}

    \KwIn{Base training data ($X_{current}$, $Y_{current}$) \\ \& Memory size (K) }
    \KwOut{Trained model parameters ($\theta_{base}$)  \\ \& Exemplar data points ($X^{[K]}_{current}$, $Y^{[K]}_{current}$)}

    $X_{train} \leftarrow X_{current}$ \\
    $Y_{train} \leftarrow Y_{current}$ \\
    $\theta_{base} \leftarrow \train(X_{train}, Y_{train})$ \\
    $Y_{predicted} \leftarrow \inference(X_{train}, Y_{train} ; \theta_{base})$ \\
    $Index_{Ranked} \leftarrow \rank(abs(Y_{predicted} - Y_{train}))$ \\
    $X^{[K]}_{current}$, $Y^{[K]}_{current} \leftarrow \retrieveExemplar(Index_{Ranked}, K)$ \\

    \KwRet{($X^{[K]}_{current}$, $Y^{[K]}_{current}$ ; $\theta_{base}$)}
    \caption{IcRegress Base Model Training}\label{alg:icregressbasemodel}
\end{algorithm}

\begin{algorithm}[t]
    \SetKwFunction{train}{train}
    \SetKwFunction{finetune}{finetune}
    \SetKwFunction{concatenate}{concatenate}
    \SetKwFunction{isAvailable}{isAvailable}
    \SetKwInOut{KwIn}{Input}
    \SetKwInOut{KwOut}{Output}

    \KwIn{Exemplar data ($X^{[K]}_{current}$, $Y^{[K]}_{current}$) \\ 
    \& New data stream ($X_{new}$, $Y_{new}$) // For adaptation \\ 
    \& Base model parameters ($\theta_{base}$)}
    
    \KwOut{Adapted model parameters ($\theta_{adapted}$)}
    $X_{train} \leftarrow X_{current}$ \\
    $Y_{train} \leftarrow Y_{current}$ \\
    
    \While{$X_{new}\neq None$}{
    $X_{train} \leftarrow \concatenate(X_{train},X_{new})$ \\
    $Y_{train} \leftarrow \concatenate(Y_{train},Y_{new})$ \\
    }

    \eIf{$\theta_{base}\neq None$}{
    $\theta_{adapted} \leftarrow \finetune(X_{train}, Y_{train} ; \theta_{base})$ \\
    }{
    $\theta_{adapted} \leftarrow \train(X_{train}, Y_{train})$ \hspace{0.5cm} // Variant (1) \\
    }
    
    \KwRet{$\theta_{adapted}$}
    \caption{IcRegress Adapted Model Training}\label{alg:icregressadaptedmodel}
\end{algorithm}

\subsection{The IcRegress Algorithm}

The IcRegress algorithm is an end-to-end incremental learning algorithm that simultaneously learns the representation of features and regressors from the data. The algorithm follows a similar structure to iCarl~\cite{rebuffi2017icarl} while adjusting the exemplar selection process and the loss function to suit a regression problem. More specifically, the exemplar data points are selected by ranking all the data points in ascending order in terms of deviating error from the ground truth rather than calculating a mean for each class as in the iCarl algorithm. Then, based on memory size, K data points are stored for continual learning and further model adaptation. Next, new data points are concatenated with K exemplar data points as the new input to fine-tune the learning algorithm and update the model weights (i.e., parameters) of the previously trained model (i.e., the pre-trained \textit{Base Model}). Algorithm~\ref{alg:icregressbasemodel} and algorithm~\ref{alg:icregressadaptedmodel} highlight the IcRegress training methodology as a two-step training approach. Algorithm~\ref{alg:icregressbasemodel} shows the \textit{Base Model} training, which outputs the parameters of the trained model and the exemplar set, while~\autoref{alg:icregressadaptedmodel} shows the incremental learning approach that produces the adapted trained model. The previously discussed back-end CNN model is used for learning in both training phases (that is, \textit{ Base Model} and \textit{Incremental Learning} training). However, the algorithm is model-invariant, and different learning models can be utilized. Furthermore, the adapted model could have a different back-end learning model for the adapted model from the base model in case of model availability issues or additional memory constraints (see \textit{Variant (1)} in algorithm~\ref{alg:icregressadaptedmodel}). However, this means the model would have to be trained longer as it is learning with random initialization instead of the pre-trained base model parameters.

\section{Results and Discussion}

The results and the discussion are divided into four sections. The first section defines the performance metrics to measure the results. The second section highlights the different results obtained through the ``Base Model'' only (without incremental learning) to compare the impact of different modalities in an ablation study. The third section highlights the results of the incremental learning approach (IcRegress); it shows the performance enhancement for the adapted model for various driver groups (e.g., amateur vs. expert drivers) and the personalization aspects. Finally, the fourth section further discusses our user-centered adaptation approach for object referencing.

\subsection{Performance Metrics}

Using State-of-the-Art (SOTA) methods~\cite{gomaa2021ml,aftab2021multimodal} to reference the object outside the vehicle, we use the Mean Square Error (MSE) as a loss function and the mean absolute error (MAE) as an error calculation metric. Opting for MAE directly provides the error in degrees, making it easily interpretable as the angular deviation from the geometric center of the target building, i.e., the ground truth. The ground truth is measured at the onset time of the referencing action as mentioned in Section~\ref{sec:baselearningmodel}. However, while the onset times have been calculated by pointing gesture recognition in~\cite{gomaa2021ml} and by a wizard-of-oz technique using a button press in~\cite{aftab2021multimodal}, we chose to calculate them using a speech command captured through a standard microphone in a more realistic setting. The speech command is filtered based on pitch and loudness to generate a binary feature that signifies whether a command occurs at a given timestamp. To demonstrate the method's superior performance, it is compared to the onset calculation using gesture recognition pointing (as in~\cite{gomaa2021ml}). Additionally, using the base model trained in an ideal setting, we demonstrate how the speech command can be employed for incremental learning when speech commands become unavailable due to a noisy driving setting.

Given the core role of performance metrics in assessing the efficacy of our proposed method and previous work on object referencing, a closer look at the metrics is warranted. Here, we deconstruct the utilization and nuances of three critical metrics in our evaluation, comparing our work with previous SOTA methods

\begin{itemize}
    \item \textbf{``Mean Relative Distance-agnostic Error'' (\textit{MRDE}):} MRDE, introduced by Gomaa et al.~\cite{gomaa2021ml}, determines whether the predicted angle lies within the geometric boundaries of the object. However, it does not account for object occlusion and the visible area of the driver, limiting its practical usage.
    
    \item \textbf{``Minimum Distance to Target'' (\textit{MinDT}):} Aftab et al.~\cite{aftab2021multimodal} proposed a method based on calculating the Euclidean distance between the reference vector and the ground truth vector. The building with the minimum distance out of five is selected as the referenced PoI. Nevertheless, this method exhibits challenges with scalability and generalization for many objects in a given surrounding
    
    \item \textbf{``Segmented Object'' (\textit{SegObj}):} Thus, we propose SegObj, a more realistic and feasible metric that considers only the visible area of the PoI (i.e., the building's facade). SegObj determines whether the predicted vector lies within the 2D bounded area of the referenced building while considering occlusion, unlike MRDE, which ignored occluded areas, and MinDT, which extended the referenced building beyond its physical space in its 2D image. SegObj is particularly more realistic, as it is compatible with current object detection algorithms, such as YOLO~\cite{bochkovskiy2020yolov4}, which return a bounding box of only the visible part of the object and overcomes the limitations of the previously mentioned metrics.
\end{itemize}

In our study, we evaluate the effectiveness of the IcRegress algorithm using these three metrics, SegObj, MRDE, and MinDT, to compare our work with similar previous studies. 
Moreover, while regression problems do not usually have a specific random chance value, we are calculating a pseudo-random chance level as a baseline for comparison. This chance level is defined as the ratio of the (PoI) angular width at the ground truth timestamp to 180 degrees. However, the method for calculating this width differs for each metric as follows.
\begin{itemize}
  \item For the MRDE, we consider the overall geometric width of the building, including any occlusion.
  \item With the SegObj, the width encapsulates only the visible part of the building's facade (i.e., the non-occluded area).
  \item In the case of MinDT, the width includes the visible facade of the building and half the air gap between adjacent buildings.
\end{itemize}
Based on these calculations, we found that the pseudo-random chance for MRDE, SegObj, and MinDT translates to $10.605\%$, $6.47\%$, and $14.45\%$, respectively. 
This suggests that our SegObj metric creates a more realistic and demanding evaluation for the task of object referencing that is more generalizable and scalable.

Finally, in order to demonstrate the performance efficacy of our incremental learning approach, IcRegress, we compare it with two traditional training approaches as a baseline, namely the ``\textit{Base Model only}'' (one-time trained model) and the ``\textit{Transfer Learning}''~\cite{torrey2010transfer} (naive fine-tuning). 
In the \textit{Base Model} approach, the model undergoes a one-time training phase utilizing the current data set. Subsequent data influx serves solely to predict without any additional training or adaptation of the existing model.
On the other hand, the alternative approach known as Transfer Learning continues the training procedure on the newly obtained data, initializing with the model previously trained with the Base Model strategy.
It should be noted that the \textit{Transfer Learning} approach can be perceived as a specific instance of our incremental learning approach when the K parameter is zero. This implies the absence of data from the previously trained model during the learning process. This technique exhibits the drawback of the ``Catastrophic Forgetting'' phenomenon, as discussed earlier in related work and observed in our results.

\subsection{Base Model and Modalities Effect}

This section presents the results of training the \textit{Base Model} (used for adaptation in the IcRegress algorithm). We evaluate our findings using the previously defined metrics: MRDE, SegObj, and MinDT. The \textit{Base Model} represents an ideal system configuration with all modalities (pointing (P), gaze (G), GazeHead (GH), and head pose (H)) in operation. The object referencing accuracy achieved is $72.4\%$, $38.2\%$, and $42.9\%$ for MRDE, SegObj, and MinDT, respectively. These values are significantly higher than the random chance of $10.605\%$, $6.47\%$, and $14.45\%$ for the same metrics.

\begin{figure}[b]
\centering
     \begin{subfigure}{0.8\linewidth}
         \centering
         \includegraphics[width=\textwidth]{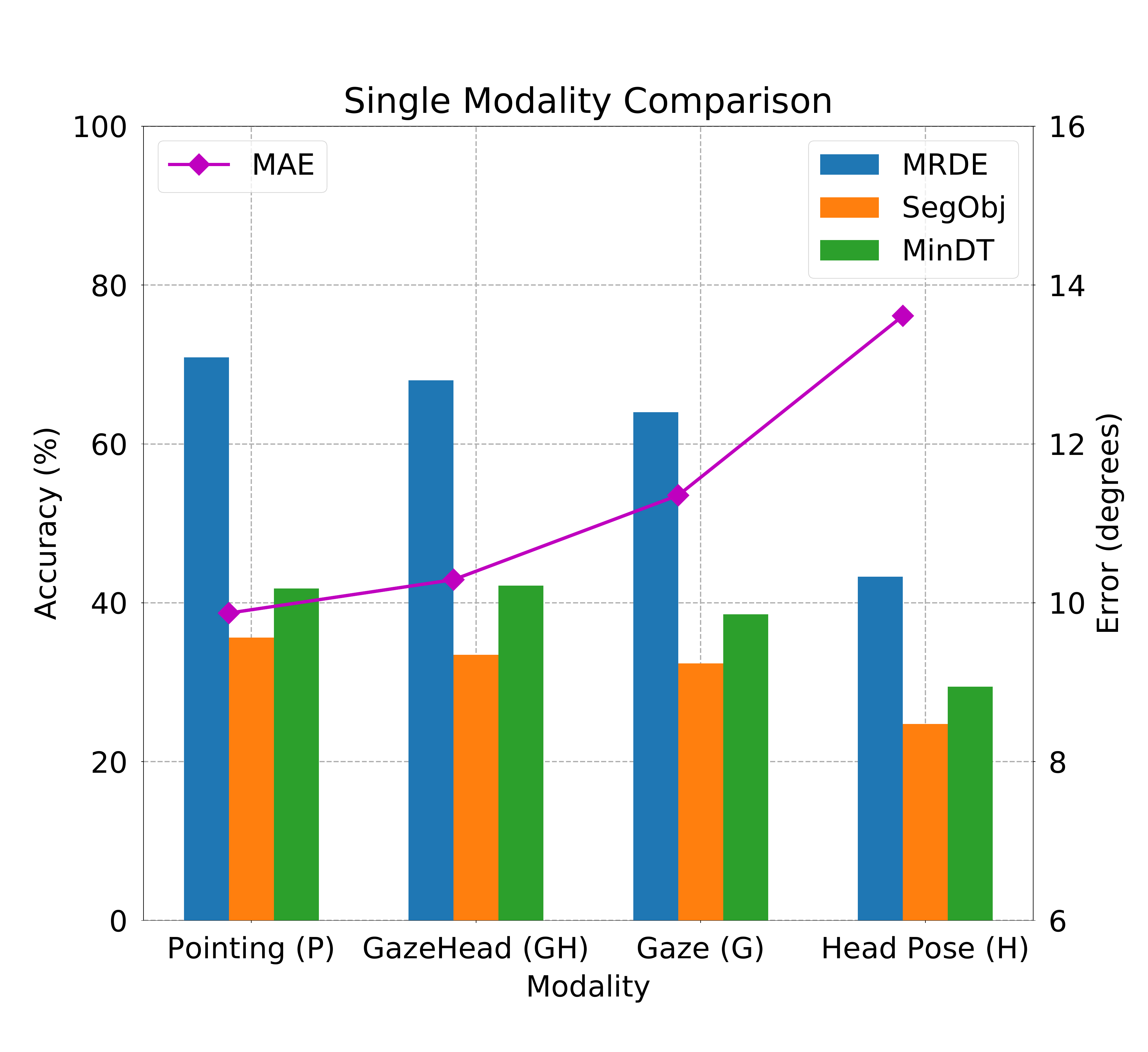}
     \end{subfigure}
     \begin{subfigure}{0.8\linewidth}
         \centering
         \includegraphics[width=\textwidth]{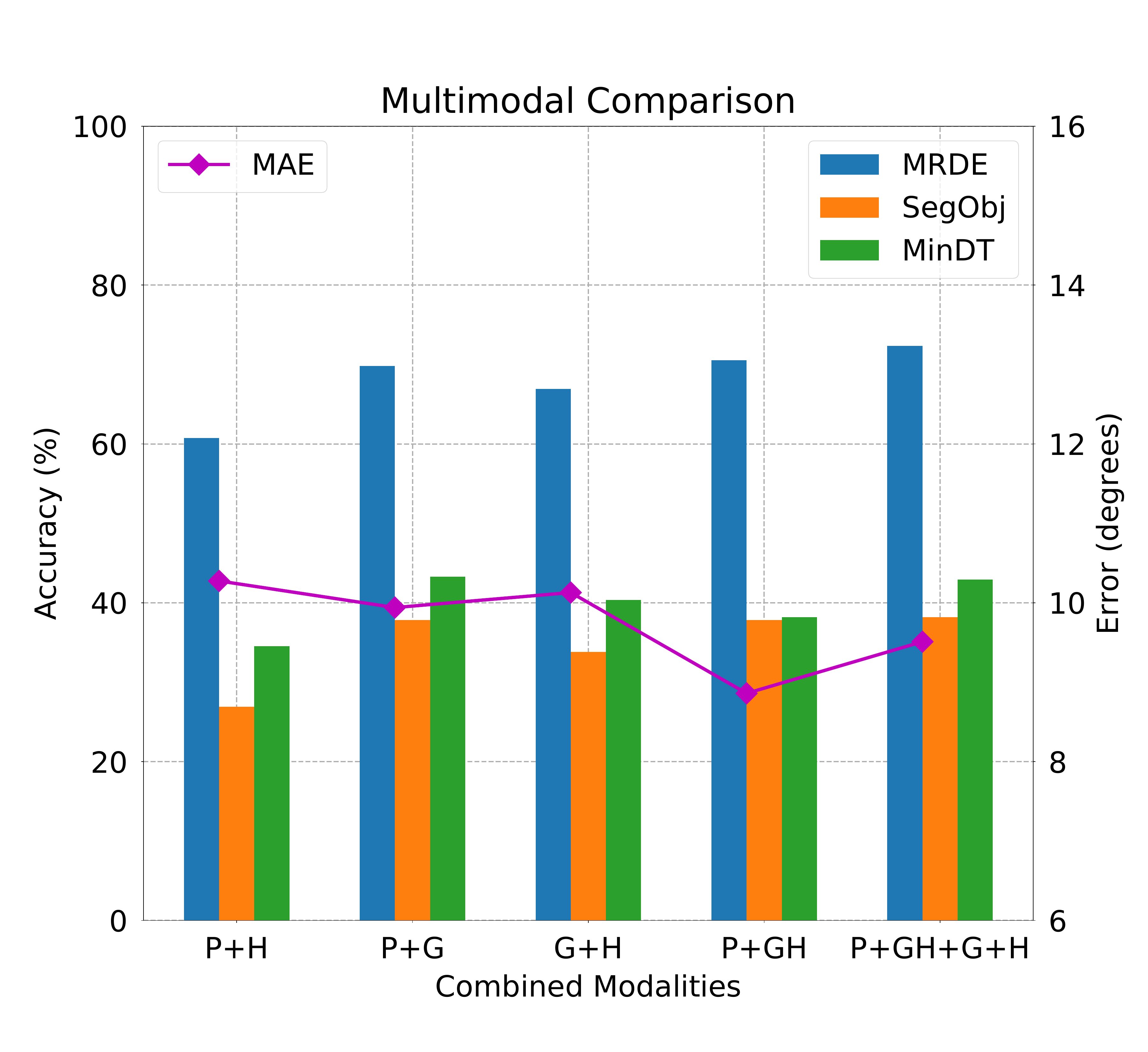}
     \end{subfigure}
    \caption{Accuracy and error results comparing single and multiple modalities performance during the multimodal object referencing task.}
     \Description{There are two bar charts in which the upper one compares single modality approaches (pointing vs. gaze vs. GazeHead vs. head pose) in terms of MRDE, SegObj, and MinDT accuracy metrics on the primary axis and MAE metric on the secondary axis (as line plot). The lower chart compares the combination of these modalities. More specifically, it has five bar charts that compare (from left to right) the combination of pointing and gaze, pointing and head pose, gaze and head pose, pointing and GazeHead, and finally, all modalities (i.e., pointing, head pose, GazeHead, and gaze).}
     \label{fig:modalitiescomparison}
\end{figure}

To better understand the significance of each modality in the learning model, we conducted an ablation study to examine each modality individually and in combination with others. \autoref{fig:modalitiescomparison} provides a comparative analysis of the individual and combined modalities.
In the individual modality analysis, pointing (P) emerged as the most effective modality for object referencing, with the highest accuracy across all metrics ($70.9\%$, $35.6\%$, and $41.8\%$ for MRDE, SegObj, and MinDT, respectively) and the lowest error rate ($9.87$ degrees for MAE). Although this aligns with previous studies by Gomaa et al.~\cite{Gomaa2020,gomaa2021ml}, which identified pointing as the dominant modality, it is not aligned with the study by Aftab et al.~\cite{aftab2021multimodal} which hypothesized that gaze is the dominant modality, as discussed next.
Regarding the gaze modality (G), an accuracy of $64\%$, $32.4\%$, and $38.5\%$ was achieved for MRDE, SegObj, and MinDT, respectively. On the other hand, the head pose (H) modality recorded lower accuracy levels of $43.3\%$, $24.7\%$, and $29.5\%$ for MRDE, SegObj, and MinDT, respectively. This is consistent with~\cite{aftab2021multimodal}'s findings that head pose is the least accurate modality. However, our results contradict the hypothesis of the same study that gaze is the dominant modality (as mentioned earlier). This discrepancy could be attributed to differences in tracking methodologies for pointing and gaze gestures or the stationary vehicle setting in the study by Aftab et al.~\cite{aftab2021multimodal}, which could have influenced driver performance.
Regarding the GazeHead (GH) modality, which combines gaze and head pose vectors, it showed improved accuracy over treating gaze and head pose as separate modalities, as suggested by~\cite{gomaa2021ml,kang2015you}. It achieved an accuracy of $68\%$, $33.5\%$, and $42.2\%$ for MRDE, SegObj, and MinDT, respectively. However,  the results were similar compared to fused gaze and head modality (refer to ``G+H'' in~\autoref{fig:modalitiescomparison}). We attribute this to the ability of the machine learning model to learn this correlation without the need for the manual preprocessing step (i.e., the arithmetic summation of the two modalities).

In our final analysis of the multimodal combination, we drew upon our previous observations. Given that pointing (P) was the dominant modality, we combined it separately with each other modality. We compared these combinations with the performance of gaze and head pose alone and the combination of all modalities (i.e., the \textit{Base Model} setting). As shown in~\autoref{fig:modalitiescomparison}, the combination of head pose and pointing (P$+$H) yielded the lowest performance ($60.7\%$, $26.9\%$ and $34.5\%$ for MRDE, SegObj, and MinDT, respectively). On the contrary, the combination of gaze and pointing (P$+$G) achieved performance metrics closest to those of the all-modality scenario (P$+$GH$+$G$+$H). This suggests that drivers do not significantly move their heads when referencing objects while driving to maintain focus on the road. This crucial finding could have significant implications for the fusion model and the referencing tracking system when used for modality switching, as discussed in previous work.
Furthermore, when comparing the combination of pointing and head pose (P$+$H) with pointing alone, we found that adding head pose degrades performance if the gaze modality does not accompany it. This is also observed in the gaze and head pose combination (G$+$H), which yields better accuracy results than each of the gaze (G) and head pose (P) modalities separately. Lastly, while the combination of pointing and GazeHead (P$+$GH) yields results comparable to those of pointing and gaze (P$+$G), it further confirms that the additional gaze head modality (GH) is unnecessary. Thus, a combination of pointing, gaze, and head pose is sufficient, contrary to suggestions made in previous work~\cite{gomaa2021ml,kang2015you}.

\subsection{IcRegress Results}

\begin{figure}[t]
\centering
     \begin{subfigure}{0.495\linewidth}
         \centering
         \includegraphics[width=\textwidth]{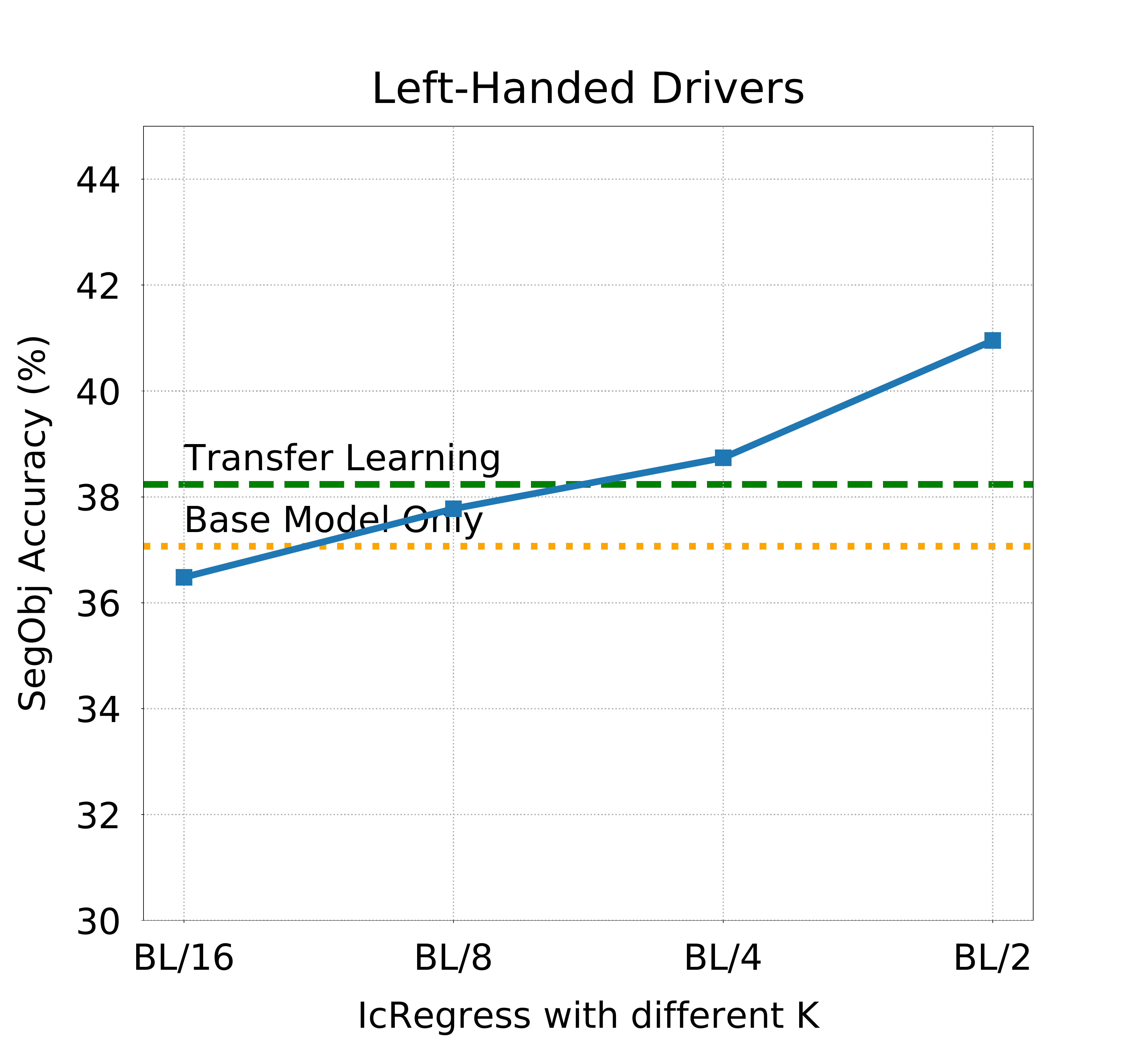}
     \end{subfigure}
     \begin{subfigure}{0.495\linewidth}
         \centering
         \includegraphics[width=\textwidth]{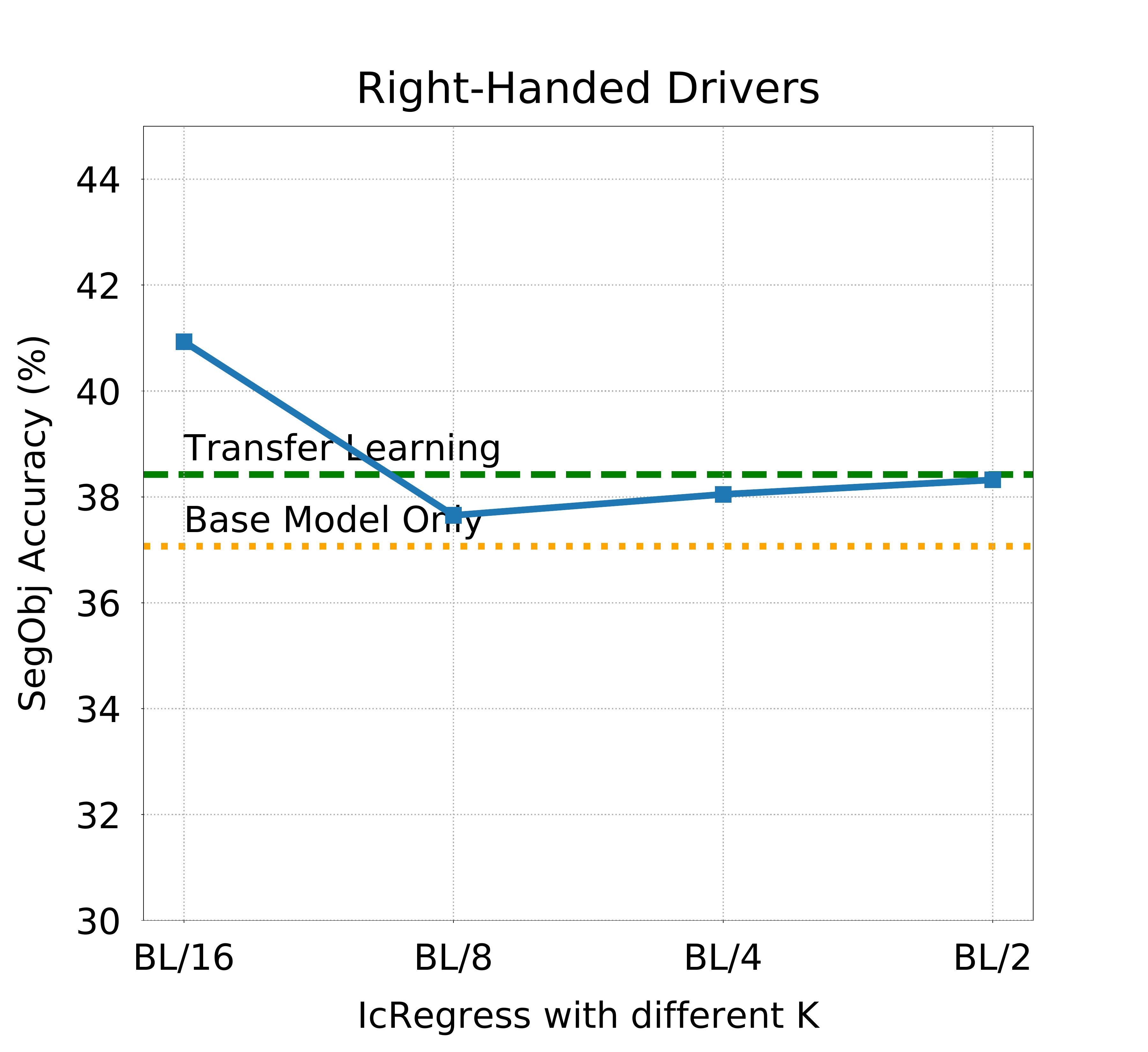}
     \end{subfigure}
    \begin{subfigure}{0.495\linewidth}
         \centering
         \includegraphics[width=\textwidth]{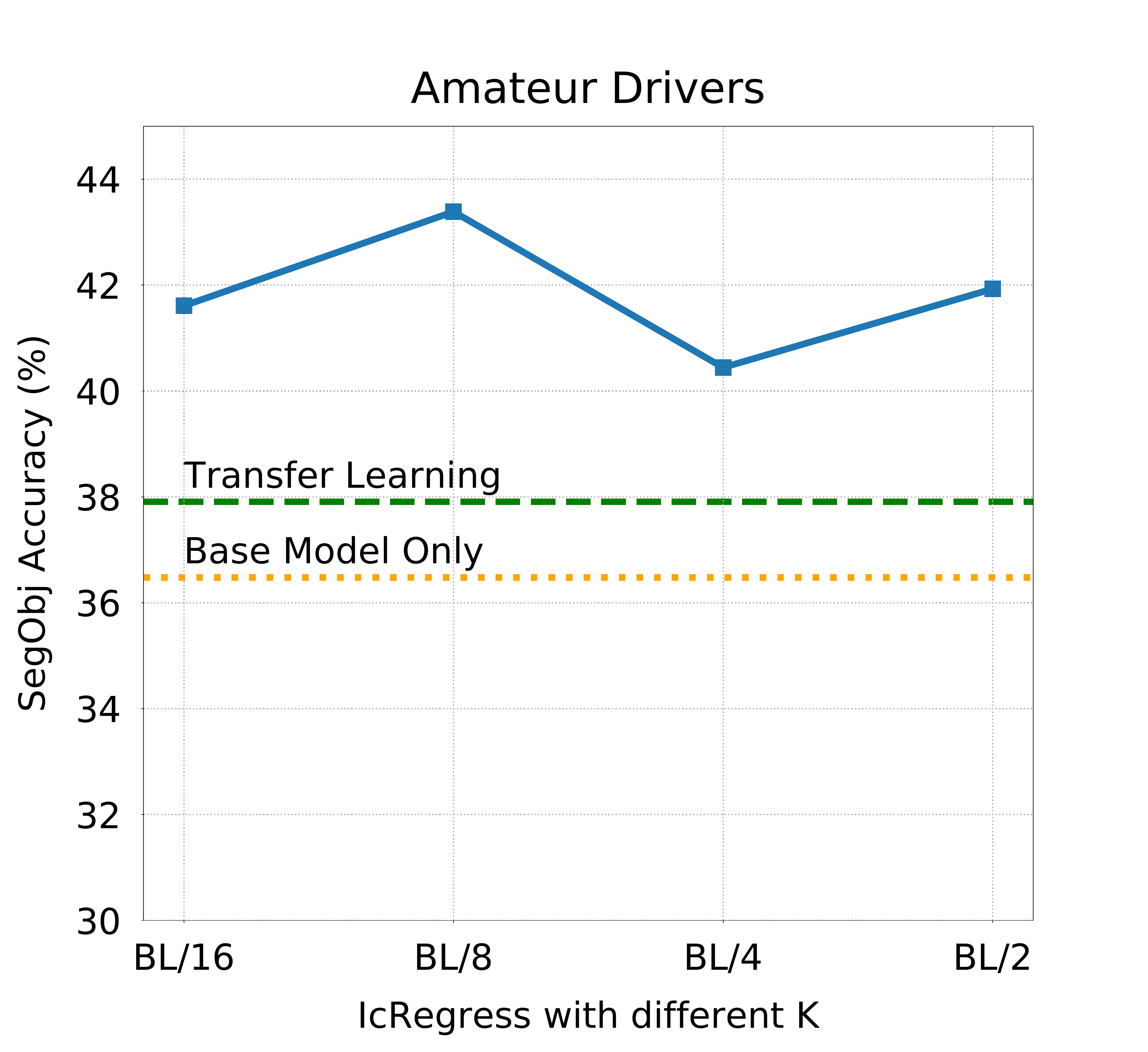}
     \end{subfigure}
     \begin{subfigure}{0.495\linewidth}
         \centering
         \includegraphics[width=\textwidth]{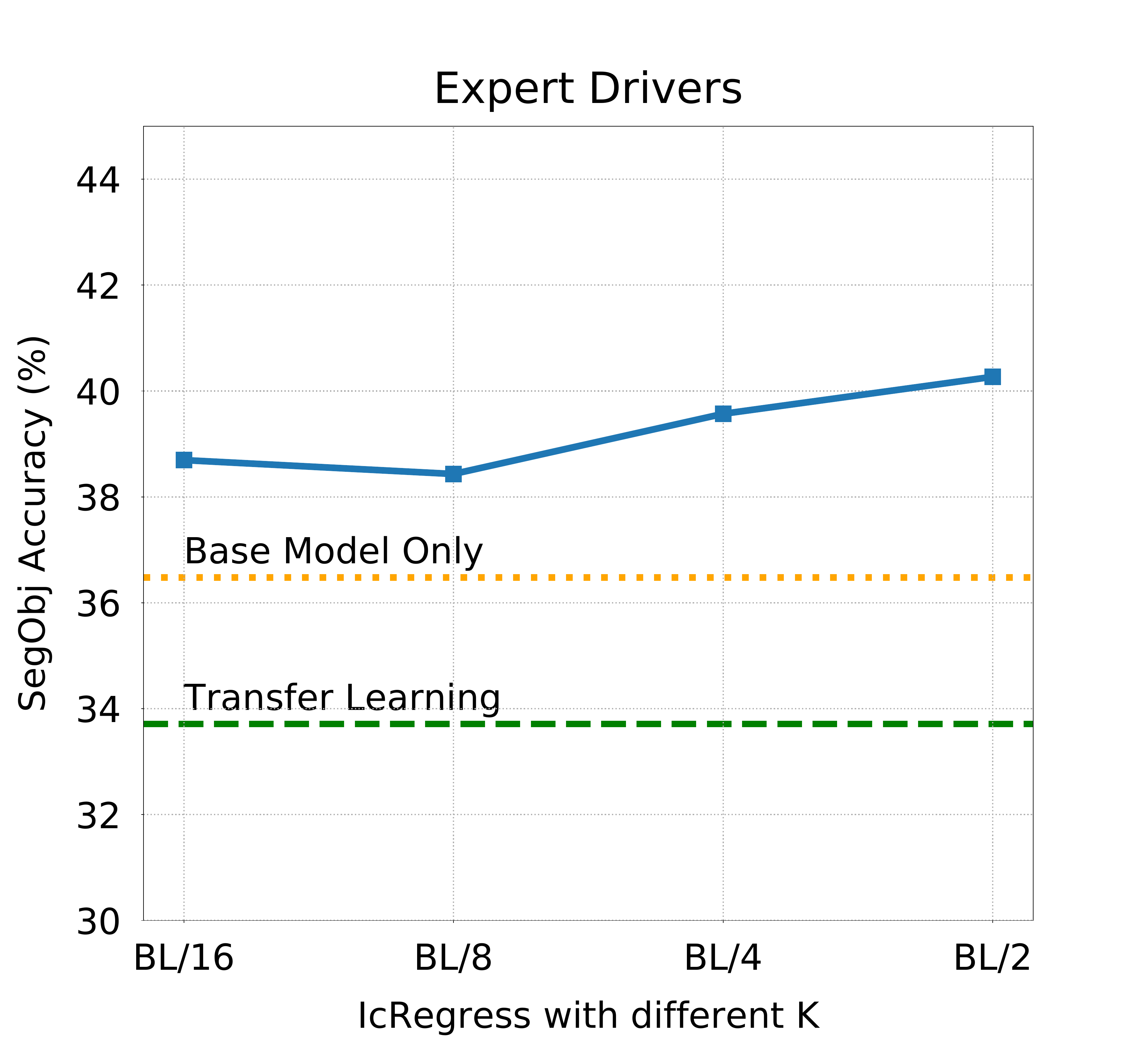}
     \end{subfigure}
     \caption{IcRegress algorithm performance regarding SegObj accuracy for drivers' different traits (i.e., handedness and driving experience). The K value is varied as a percentage from the base model length (BL) by selecting one-sixth of the BL to half of the BL and using it in the adapted model. IcRegress is compared against an inference-only approach (i.e., the base model only) and the Transfer Learning approach.}
     \Description{There are four line charts comparing the IcRegress algorithm with different K-values to two baselines (indicated as horizontal lines), an inference (i.e., base model only) and a transfer learning baseline. The charts are, from left to right, left-handed drivers, right-handed drivers, amateur drivers, and expert drivers. It can be seen that the IcRegress accuracy values are higher than the base model only for all k-values except K=BL/16.}
     \label{fig:Icregress1}
\end{figure}

After training the \textit{Base Model}, we utilize it for incremental learning using the IcRegress algorithm. A crucial hyperparameter for most incremental learning algorithms, including IcRegress, is the proportion of old training data (i.e., \textit{Base Model} input data) to incorporate when fine-tuning the new training data (i.e., the adapted model). We adjust this hyperparameter (denoted as K) for the old data percentage, ranging from one-sixteenth to half. According to incremental learning theory and previous work, the accuracy numbers (i.e., enhancement gain) should increase as the K-value increases. 
To demonstrate the performance of IcRegress, we compare it with the two previously described baselines. The first baseline is the \textit{Base Model only} (one-time trained model), where the \textit{Base Model} is not further trained and is applied to the new data in inference mode. The second baseline employs \textit{Transfer Learning} (naive fine-tuning), where only the parameters (i.e., weights and biases of the trained model) of the \textit{Base Model} are used for further training on new data without including any of the old data points. This is equivalent to an IcRegress learning model with K equal to zero. 

\begin{figure}[t]
\centering
     \begin{subfigure}{0.495\linewidth}
         \centering
         \includegraphics[width=\textwidth]{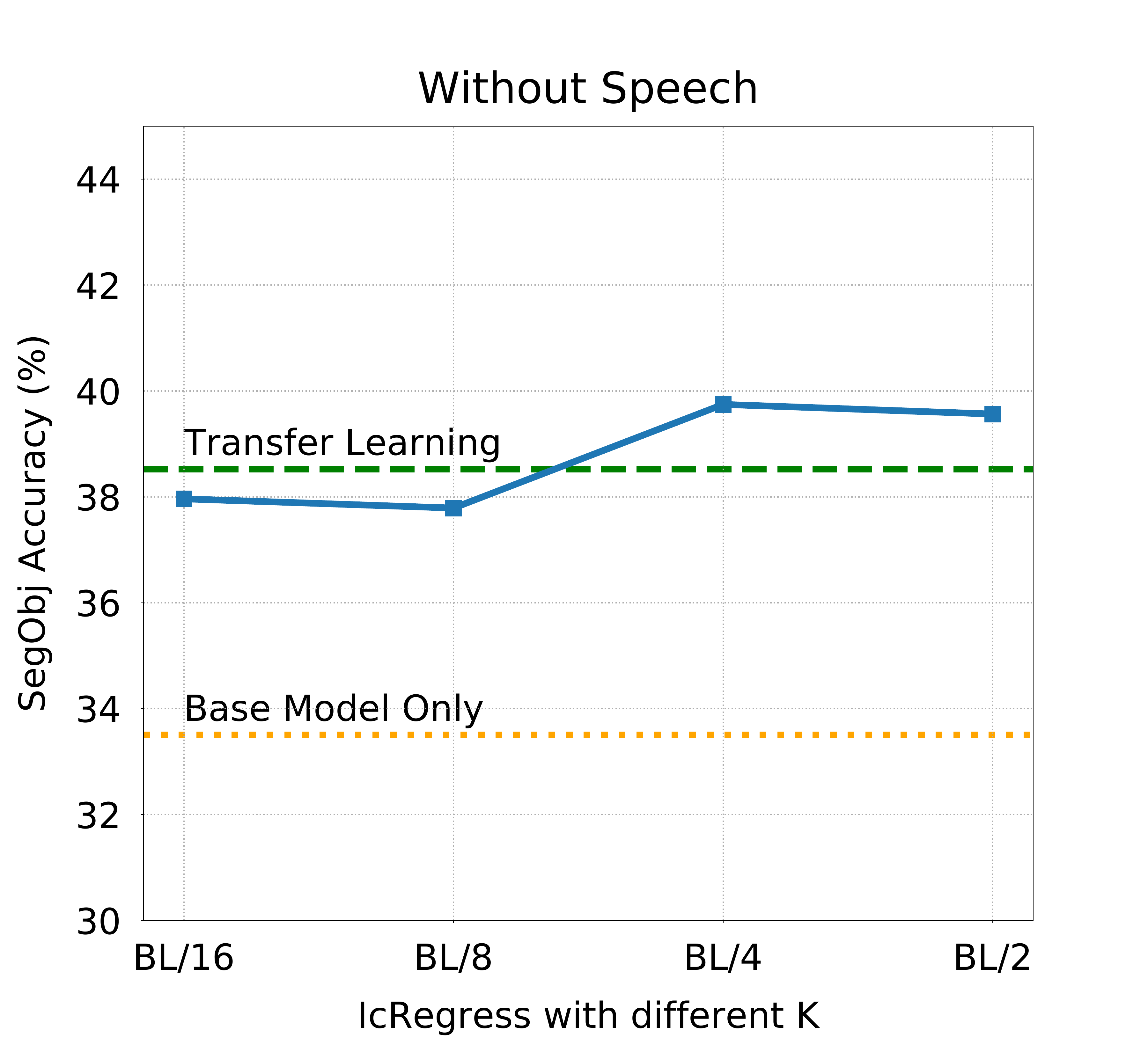}
     \end{subfigure}
     \begin{subfigure}{0.495\linewidth}
         \centering
         \includegraphics[width=\textwidth]{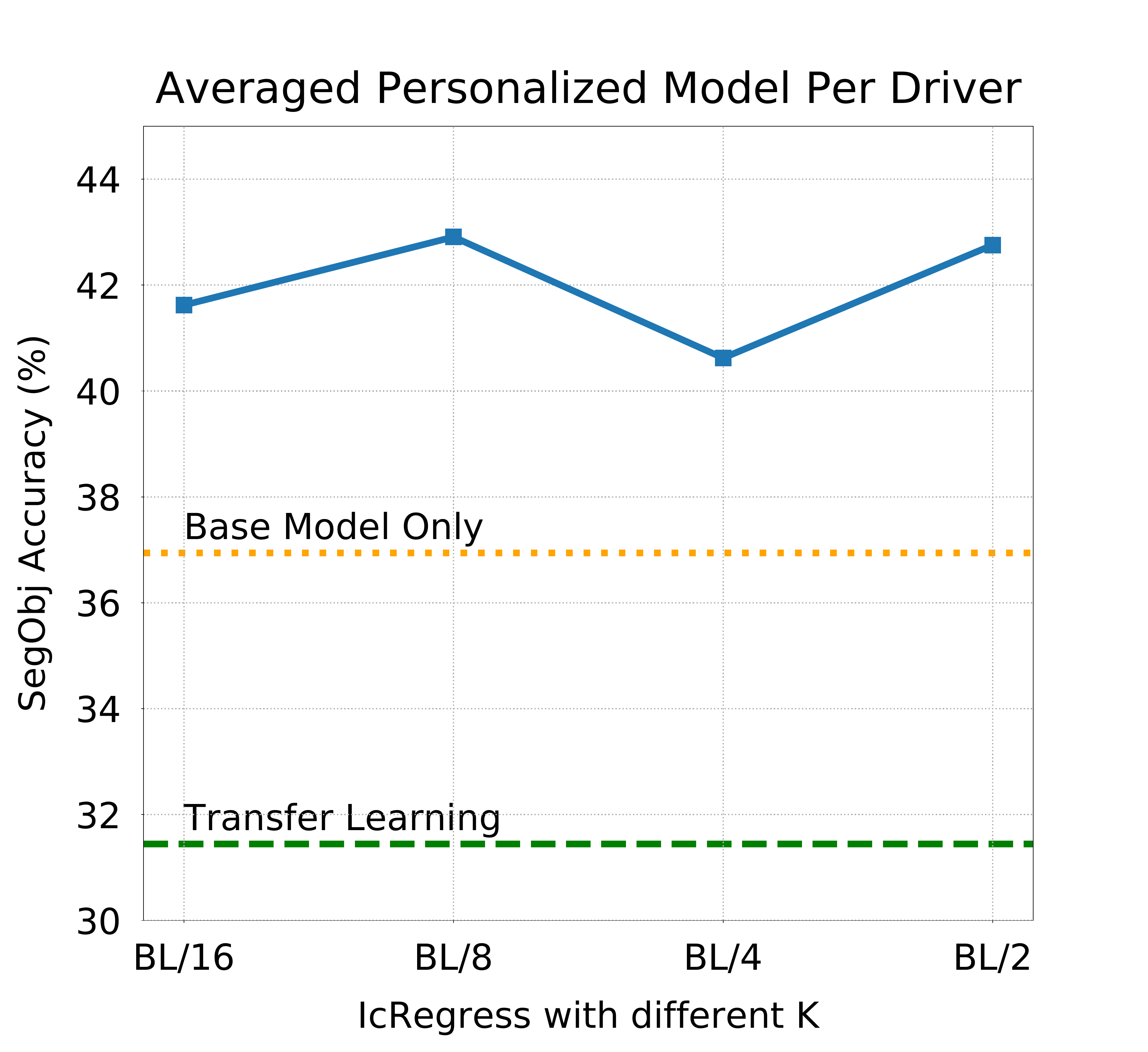}
     \end{subfigure}
     \caption{IcRegress algorithm performance in terms of SegObj accuracy for personalized adaptation (right) and the lack of speech command setting (left). The K value is varied as a percentage from the \textit{Base Model Length} (BL) by selecting one-sixteenth of the BL to half of the BL and using it in the adapted model. IcRegress is compared against an inference-only approach (i.e., the \textit{Base Model only}) and the \textit{Transfer Learning} approach.}
     \Description{There are two line charts comparing the IcRegress algorithm with different K-values to two baselines (indicated as horizontal lines), an inference (i.e., \textit{Base Model only}) and a \textit{Transfer Learning} baseline. The graphs show the performance of the data without speech commands on the left and the average of personalized models per driver. It can be seen that the IcRegress accuracy values are higher than the \textit{base model only} for all k-values except K=BL/16.}
     \label{fig:Icregress2}
\end{figure}

\autoref{fig:Icregress1} and~\autoref{fig:Icregress2} present the IcRegress SegObj accuracy results for different K values for various driver traits compared to the \textit{Base Model only} and \textit{Transfer Learning} baselines. For all different driver traits and personalized models, the IcRegress algorithm surpasses the \textit{Base Model only} for almost all K values. However, it does not outperform the \textit{Transfer Learning} approach for all K values. 
More specifically, in~\autoref{fig:Icregress1}, IcRegress surpasses the \textit{Transfer Learning} baseline for all K values for the driving experience (i.e., amateur vs. experienced drivers). However, it only outperforms \textit{Transfer Learning} for left-handed drivers for K values of at least 25\% percent of the \textit{Base Model} data and only outperforms \textit{Transfer Learning} for right-handed drivers for K values greater than half of the \textit{Base Model} data. Similarly, for the missing speech command and the personalized drivers in~\autoref{fig:Icregress2}, IcRegress outperforms the \textit{Transfer Learning} baseline for all K values for personalized models. However, it outperforms \textit{Transfer Learning} for data without speech command only for K values greater than 25\% of the base data.

These figures also illustrate that the issue of ``Catastrophic Forgetting'' is not severe in instances where \textit{Transfer Learning} outperforms the \textit{Base Model only}. However, it becomes significant in the case of expert drivers and personalized models. In these instances, the models are entirely overfitted to specific driver groups, resulting in performance inferior to the \textit{Base Model only}. This observation confirms the existence of ``Catastrophic Forgetting'' in the trained models, an issue IcRegress has successfully addressed. We analyze the personalized model adapted for each driver to investigate these results further. 
Personalized models are developed by splitting the test participants' data into two halves: the training and the test half. The training half is treated as the new data stream for adaptation, while the test half is used for model evaluation. Consequently, \autoref{fig:Icregress2} shows each personalized model's average test set accuracy. For further analysis, we divided the test set into two subsets. One subset was sampled from the same distribution as the adapted model (i.e., the current driver), and the other was sampled from a different distribution of non-adapted participants (i.e., other ``possible'' drivers) on which the model was not trained. This division strategy was employed in previous studies~\cite{gomaa2021ml,gomaa2022adaptive} to evaluate the effectiveness of personalization. 
In addition, we trained a model from scratch using only the adapted participant training data to underscore the overfitting effect further and make it more apparent..

\begin{figure}[t]
	\begin{center}
		\includegraphics[width=\linewidth]{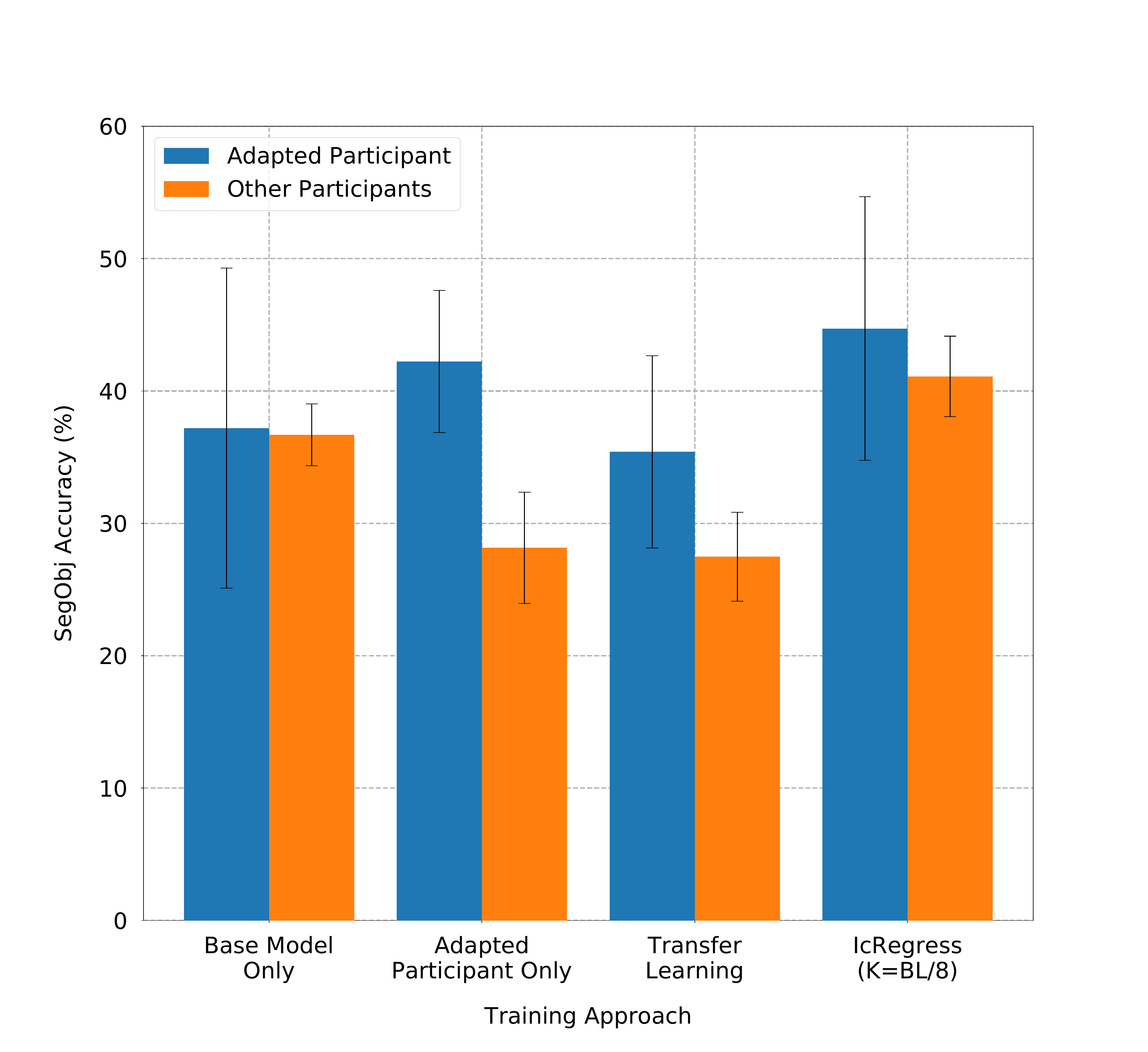}
	\end{center}
     \caption{Comparison of performance among inference using \textit{Base Model only}, \textit{Transfer Learning}, retraining from scratch on the new participant's data, and IcRegress with K equal one-eighth of the base model length approaches. The results are obtained for two test sets, one from the same adapted participant and one from the other non-adapted participants. The chart shows the confidence interval over all test participants.}
     \Description{This is a bar chart with four ticks on the x-axis and two bars per tick showing adapted vs. other participants. On the y-axis is the SegObj accuracy. The base model-only approach has the same accuracy value for both the adapted and other participants around $37\%$, which IcRegress exceeds with values of $45\%$ and $41\%$. The adapted participant has higher values than other participants, with a noticeable gap for the adapted participant only and transfer learning cases.}
	\label{fig:personalization}
\end{figure}

\autoref{fig:personalization} presents the results of the previously described analysis. It compares the two previously discussed baselines, the training from scratch on new data only and the IcRegress approach with K set to one-eighth of the base data (this K-value is chosen as a trade-off between performance and memory usage). Given that these data are averaged across participants, we have included a confidence interval for the accuracy value between test participants. 
The results indicate that both the training on new data only and the \textit{Transfer Learning} approach attempt to improve the \textit{Base Model} by overfitting on the adapted participant. However, these methods do not generalize well to other participants, leading to a discrepancy in the accuracy of performance results. This observation is consistent with suggestions and findings from previous research, which recommends training separate adapted models for personalization without incremental learning~\cite{gomaa2021ml}. On the other hand, the IcRegress algorithm enhances both the same distribution (i.e., the current user) and different distribution (i.e., other users) test sets by incorporating a representative of the old distribution, demonstrating superior generalization capabilities and overcoming ``Catastrophic Forgetting''.

\subsection{User-centered Design for Object Referencing}

Despite significant technological advancements in the automotive sector, driving remains a highly complex task that can cause mental overload~\cite{groeger2000understanding, urry2016mobilities}. Consequently, drivers interact with vehicular interfaces in various ways while driving. Particularly for the object referencing task, drivers' performance can fluctuate based on various factors such as their driving experience, dominant hand, and current emotional state, including feeling relaxed or restrained. Therefore, to achieve high accuracy in object referencing, it is imperative to adapt to different drivers and the same driver under different circumstances (see~\autoref{fig:teaser} for examples). 
Incremental learning approaches, such as IcRegress, are well suited to adapting to drivers' changing behavior and unique characteristics during the object referencing task. IcRegress can modify the model distribution (i.e., model parameters) to align with new situations over time without discarding trained information from past situations by continually adjusting the trained machine-learning model. Furthermore, it automatically accommodates temporary changes in the driver's behavior due to sudden environmental shifts. In essence, this adapted approach considers implicit feedback from the driver in a user-centered manner, thereby fostering trust and enhancing the user experience.

\section{Limitation and Future Work}

Although the IcRegress algorithm outperforms the performance of the \textit{Base Model only} and \textit{Transfer Learning} approaches, there are certain limitations that we aim to address in future work. First, the dataset utilized in this study is relatively small. Therefore, it would be beneficial to test the algorithm with a larger dataset encompassing a more comprehensive range of driving scenarios, such as increased traffic flow or the introduction of external distractions, such as a talking passenger. Furthermore, while a simulated environment offers better control over the design of these driving scenarios, real-world driving situations could present additional factors to consider. As such, future work could involve testing the IcRegress algorithm in a real-world setting. 

Moreover, the proposed model adapts to the driver's behavior over time, but there may be instances where real-time adaptation is required. In future work, we plan to explore the real-time adaptation of the model by integrating incremental learning techniques with online learning techniques, where data are available only as a stream of single data points. Lastly, while the proposed model focuses primarily on the technical aspects of the interaction between the driver and the vehicle, it is crucial to consider human factors such as the driver's cognitive load, attention, and situational awareness. Although these factors are implicitly considered in driver traits (e.g., driving experience), we aim to incorporate them into future work explicitly.

\section{Conclusion}

Traditional machine learning approaches typically involve training models on fixed datasets, which do not account for changes in the underlying data distribution over time. In contrast, incremental learning algorithms enable models to adapt to new data over time, thereby enhancing the model's accuracy and robustness. Within the context of object referencing, drivers exhibit variations in gestural input, a challenge that can be mitigated with large data-driven machine-learning models. However, incremental learning techniques can capture these differences and adapt the model to the driver's unique characteristics and behavior without relying on massively large datasets. This results in a more personalized and adaptable approach to multimodal gestural input, improving driver safety and convenience.
Thus, in this work, we propose a novel incremental learning approach, IcRegress, to disambiguate the referenced object of the driver through the multimodal fusion of gaze, head pose, pointing, and speech commands. This approach adapts to different drivers, driving scenarios, and specific driver states. Additionally, we address the limitations of existing outside-the-vehicle object referencing approaches by introducing new practical metrics for measuring performance in a driving simulation environment. These metrics can be transferred to real-world scenarios, thus enhancing the practical applicability of our work and previous work. Additionally, we conducted an ablation study to assess the importance of each modality in performing the referencing task.
Finally, the proposed approach aims to improve user experience, acceptance, and trust in human-vehicle interaction by offering more adaptable and personalized interfaces~\cite{detjen2021increase}. We provide an open-source framework for the IcRegress algorithm, which can be utilized specifically for outside-the-vehicle object-referencing tasks and, more generally, for incremental learning regression problems. This work lays the foundation for future research in this domain and opens new avenues to enhance the interaction between humans and vehicles in a personalized and efficient manner.

\begin{acks}

This work is partially funded by the German Ministry of Education and Research (BMBF) under the FedWell project (Grant Number: 01IW23004) and the CAMELOT project (Grant Number: 01IW20008).

\end{acks}

\bibliographystyle{ACM-Reference-Format}
\bibliography{sample-base}


\end{document}